\documentclass[prx,preprint,showpacs,preprintnumbers,amsmath,amssymb,citeautoscript,superscriptaddress]{revtex4-1}
\usepackage{graphicx}
\usepackage{amssymb, amsfonts, amsmath}
\usepackage{dcolumn}
\usepackage{bm}
\usepackage{color}
\usepackage{here}

\newlength{\figwidth}
\begin{document}

\title{Nematic quantum critical point without magnetism in FeSe$_{1-x}$S$_{x}$ superconductors}

\author{S. Hosoi}
\author{K. Matsuura}
\author{K. Ishida}
\author{Hao Wang}
\author{Y. Mizukami}
\affiliation{Department of Advanced Materials Science, University of Tokyo, Kashiwa, Chiba 277-8561, Japan}
\author{T. Watashige}
\author{S. Kasahara}
\author{Y. Matsuda}
\affiliation{Department of Physics, Kyoto University, Sakyo-ku, Kyoto 606-8502, Japan}
\author{T. Shibauchi}
\affiliation{Department of Advanced Materials Science, University of Tokyo, Kashiwa, Chiba 277-8561, Japan}
\date{\today}

\begin{abstract}
{\bf
The importance of antiferromagnetic fluctuations are widely acknowledged in most unconventional superconductors \cite{Moriya2003,Scalapino2012,Hirschfeld2011}. In addition, cuprates and iron pnictides often exhibit unidirectional (nematic) electronic correlations \cite{Keimer2015,Fradkin2015,Fernandes2014}, including stripe and orbital orders, whose fluctuations may also play a key role for electron pairing \cite{Maier2014,Metlitski2015,Lederer2015,Kontani2010}. However, these nematic correlations are intertwined with antiferromagnetic or charge orders, preventing us to identify the essential role of nematic fluctuations. This calls for new materials having only nematicity without competing or coexisting orders. Here we report systematic elastoresistance measurements in FeSe$_{1-x}$S$_{x}$ superconductors, which, unlike other iron-based families, exhibit an electronic nematic order without accompanying antiferromagnetic order. We find that the nematic transition temperature decreases with sulphur content $x$, whereas the nematic fluctuations are strongly enhanced. Near $x\approx0.17$, the nematic susceptibility diverges towards absolute zero, revealing a nematic quantum critical point. This highlights FeSe$_{1-x}$S$_{x}$ as a unique nonmagnetic system suitable for studying the impact of nematicity on superconductivity.
}
\end{abstract}

\maketitle

  In underdoped cuprate superconductors, unidirectional electronic correlations (stripe correlations) appear in the pseudogap state, whose relation with superconductivity is a centre of debate. It has become more complicated after the charge density wave (CDW) order has been observed in a portion of this pseudogap region of the phase diagram \cite{Keimer2015}. In iron pnictides, the tetragonal-to-orthorhombic structural transition always precedes or coincides with the antiferromagnetic (AFM) transition \cite{Hirschfeld2011}. Below the structural transition temperature $T_{\rm s}$, electronic nematicity that represents a large electronic anisotropy breaking the $C_4$ rotational symmetry, is observed \cite{Fernandes2014}, which may have a similar aspect with the stripe correlations in underdoped cuprates. In both cases, however, the nematicity is largely coexisting and intertwined with other CDW and AFM orders. Large nematic fluctuations have been experimentally observed in BaFe$_2$As$_2$ systems above $T_{\rm s}$, and these nematic fluctuations are strikingly enhanced with approaching the end point of the structural transition \cite{Yoshizawa2012,Kuo2015}. This suggests the presence of a nematic quantum critical point (QCP), but in this case the nematic QCP coincides with, or locates very close to the AFM QCP, where antiferromagnetic fluctuations are also enhanced \cite{Shibauchi2014}. This raises a fundamental question as to which fluctuations are the main driving force of the Cooper pairing in this system.   
 
  Recently there is growing evidence that the iron-chalcogenide FeSe displays remarkable properties.  The superconducting transition temperature of $T_{\rm c}=9$\,K at ambient pressure is largely enhanced up to 38\,K under pressure \cite{Sun2015}. Further enhancement of $T_{\rm c}$ has been reported in monolayer FeSe \cite{Wang2012}. What is important in this system is that despite its high nematic transition temperature of $T_{\rm s}\approx90$\,K \cite{Boehmer2013}, no magnetic order occurs down to $T\to0$\,K. Above $T_{\rm s}$,  large nematic fluctuations are reported but, unlike iron pnictides, no sizable AFM fluctuations are observed \cite{Baek2015}. The enhanced nematic fluctuations most likely have an orbital origin, because a momentum-dependent orbital polarization has been found in angle-resolved photoemission spectroscopy in the nematic phase below $T_{\rm s}$ \cite{Suzuki2015}.  More recently, it has been shown that the high-energy spin excitations measured by the inelastic neutron scattering do not show significant temperature dependence above $T_{\rm s}$ \cite{Shamoto2015}, which indicates no direct correlations between dynamical spin susceptibility $\chi({\bf q},T)$ and the nematic susceptibility, inconsistent with the spin-nematic scenario \cite{Fernandes2014}. Although these results point to that the nematic transition in FeSe is orbital driven, the most important issue, namely the presence of a nonmagnetic nematic QCP at which the nematic fluctuation diverges with $T\to 0$\,K, remains open.  The presence or absence of such a QCP should provide pivotal information on the superconductivity of iron-based high-$T_{\rm c}$ materials.  It has been shown very recently that the applying pressure leads to the suppression of $T_{\rm s}$, but near the critical pressure at which $T_{\rm s}$ vanishes, the magnetic ordering is induced \cite{Sun2015}, preventing our understanding of the interplay between nematicity and superconductivity.

 Recent advances in the crystal growth of iron chalcogenides by using the chemical vapour transport technique \cite{Boehmer2013} allow us to grow high-quality single crystals of FeSe$_{1-x}$S$_x$.  The high quality of the obtained crystals is evidenced by the observation of quantum oscillations in a wide range of $x$ (Kasahara, S. {\it et al.}, and Coldea, A.\,I. {\it et al.}, unpublished results). Here we report that the partial substitution of isoelectric S for the Se site \cite{Mizuguchi2008,Guo2014,Watson2015,Moore2015} leads to the complete suppression of this nematic transition without inducing the AFM order.  Moreover $T_c$ shows a maximum at $x\approx 0.08$ in FeSe$_{1-x}$S$_x$, suggesting the presence of superconducting dome.  Therefore FeSe$_{1-x}$S$_x$ appears to be a key material to investigate a possible correlation between the nonmagnetic nematic fluctuations and the superconductivity.  
 
 The temperature dependence of the in-plane resistivity $\rho(T)$ in these high-quality crystals shows a kink at the nematic transition at $T_{\rm s}$, which is gradually decreased to lower temperatures (Fig.\;1a). At high concentrations $x\ge0.17$, the kink anomaly of $\rho(T)$ disappears, which is more clearly seen in the temperature derivative ${\rm d}\rho/{\rm d}T(T)$ (Fig.\;1b). In contrast to the physical pressure case where another clear anomaly at a temperature different from $T_{\rm s}$ is reported indicating the appearance of a magnetic transition, we find no such anomaly in the entire $x$ range of this study, which implies that the ground states of this system are nonmagnetic up to at least $x=0.21$. 

An elegant way to evaluate experimentally the nematic fluctuations has been developed by the Stanford group, that is based on the elastoresistance measurements by using a piezoelectric device \cite{Chu2012,Kuo2013}. We utilize this technique in our FeSe$_{1-x}$S$_x$ to extract the temperature dependence of the so-called nematic susceptibility $\chi_{\rm nem}$. Here the change in the resistance $N \sim (\Delta R/ R)$ induced by the strain that can be controlled by the voltage applied to the piezoelectric device is measured as a function of strain $\epsilon$ (Fig.\;2a). This quantity $N$ is proportional to the nematic order parameter, and thus its fluctuations can be quantified by the nematic susceptibility $\chi_{\rm nem} \sim \frac{\partial N}{\partial \epsilon}$, which is given by the slope of the $(\Delta R/ R)$ versus $\epsilon$ curve \cite{Chu2012}. To cancel the effect of anisotropic strain induced in the piezoelectric device (the Poisson effect) \cite{Kuo2013}, we measured the resistance changes $(\Delta R/ R)_{xx}$ and $(\Delta R/ R)_{yy}$ for two current directions orthogonal each other in FeSe ($x=0$) crystals (Figs.\;2a, b). The nematic susceptibility data obtained from $(\Delta R/ R)_{xx}$ and $(\Delta R/ R)_{yy}$ are essentially identical to those from $(\Delta R/ R)_{xx}$ alone (Fig.\;3a). Therefore, we measure only $(\Delta R/ R)_{xx}$ for substituted samples $x>0$ to avoid the effect of small difference of the composition $x$ in different crystals from the same batch. The sign of the nematic susceptibility above $T_{\rm s}$ is positive, which corresponds to the positive value of $\rho_{a} -\rho_{b}$ where $a$ and $b$ represent the orthorhombic crystal axes ($a>b$). This positive sign is consistent with the previous results in FeSe \cite{Watson2015,Tanatar2015} and FeTe$_{0.6}$Se$_{0.4}$ \cite{Kuo2015}, but is opposite to the Co-doped BaFe$_2$As$_2$ case \cite{Chu2012}. This sign difference may be related to the difference of Fermi surface structure, which determines the anisotropic scattering of quasiparticles in the nematic state \cite{Blomberg2013}. 

The temperature dependence of the obtained nematic susceptibility shows a systematic trend as a function of the sulphur content $x$ (Figs.\;3a-e). In the tetragonal state above $T_{\rm s}$, $\chi_{\rm nem}(T)$ can be fitted to the Curie-Weiss type temperature dependence given by  
\begin{equation}
\chi_{\rm nem} = \frac{\lambda}{a (T -T_{\theta})} + \chi^{0},
\label{eq1}
\end{equation}
where $\lambda$ and $a$ are constants, and $T_{\theta}$ is the Weiss temperature, which represents the electronic nematic transition temperature without the lattice coupling. The observed changes in the resistance is given by $\Delta R/R = \Delta \rho / \rho + \Delta L/L - \Delta A/A$, the last two terms on the right hand side represent geometric factors related to changes in the length $L$ and cross-sectional area $A$ \cite{Kuo2013}. These terms contribute to the temperature-independent term $\chi^{0}$ in Eq.\;(1), which is dominant in simple metals. In the present FeSe$_{1-x}$S$_x$ system, however, $\chi_{\rm nem}(T)$ has strong temperature dependence, indicating that nematic fluctuations are very large. With a proper choice of $\chi^{0}$, $1/(\chi_{\rm nem}-\chi^{0})$ follows nearly $T$-linear dependence in a wide temperature range of the tetragonal phase, consistent with the Curie-Weiss law (Eq.\;(1)). 

The obtained phase diagram based on our single-crystal study is given by Fig.\;4. The nematic transition temperature $T_{\rm s}$ monotonically decreases and vanishes at $x\gtrsim0.17$. The Weiss temperature $T_{\theta}$ crosses the zero line and changes sign at $x\approx0.17$, where the magnitude of $\chi_{\rm nem}$ is strongly enhanced. These results provide clear evidence that the nematic QCP is present near this composition $x=0.17$, where nematic fluctuations are diverging toward $T\to0$\,K. We stress that the present measurements are performed under identical conditions in a series of samples with similar dimensions and thus the obtained systematic results represent intrinsic properties of this system. We also note that the recent quantum oscillations and low-field magnetoresistance experiments indicate that the electronic structure of high-concentration samples ($x\sim0.21$) is quite different from that of low-$x$ crystals (Kasahara, S. {\it et al.}, and Coldea, A.\,I. {\it et al.}, unpublished results), which is indicative of distinctly different ground states in the two regimes, consistent with the quantum phase transition found in the present study. 

Near the QCP, the $\chi_{\rm nem}(T)$ data exhibit some deviations from the Curie-Weiss temperature dependence (grey regions in Figs.\;3c-e, h-j). Such deviations have also been frequently found in iron-based superconductors when $T_{\rm s}$ becomes low \cite{Kuo2015}. The origin of these deviations from the Curie-Weiss law has not been fully understood, but one should consider possible effects of impurity scattering that may be modified by quantum criticality.  In the theory of magnetic quantum critical point, various physics quantities show non-Fermi liquid behaviour, and magnetic susceptibility may deviate from the Curie-Weiss law \cite{Moriya2003}. Magnetic susceptibility directly couples to the correlation length, which depends on dimension and dynamical exponent of the system according to the Hertz-Millis theory \cite{Lohneysen2007}. In analogy to this, the nematic correlation length in the presence of disorder would have some characteristic temperature dependence different from the Fermi-liquid behaviour, and then the deviations from Curie-Weiss law may be expected in the nematic susceptibility. When the effects of disorder is strong, it has been discussed that the so-called Griffith singularity may occur, which modifies the critical behaviour \cite{Kuo2015}. In any case, these disorder effects should weaken the divergent behaviour especially near the QCP, which is consistent with our results. 

The present results provide evidence that a nonmagnetic nematic QCP lies in the superconducting phase, which has the same topology as the phase diagrams of iron pnictides \cite{Shibauchi2014} and heavy fermions \cite{Lohneysen2007} in which AFM QCP locates inside the superconducting dome. However, this new kind of QCP has several different aspects from the well-studied AFM QCP.  First, the resistivity data (Fig.\;1a) shows non-$T^2$ dependence for all the sample we measured, and we find no significant enhancement in the residual resistivity near the nematic QCP, which appears to be different from the expectations in the magnetic QCP case \cite{Lohneysen2007}.  Second, in iron pnictides $T_{\rm c}$ shows the maximum at around the AFM QCP \cite{Shibauchi2014}, wheres in the present system $T_{\rm c}$ shows a maximum deep inside the nematic ordered phase, not at the nematic QCP. This suggests that the critical nematic fluctuations do not have simple correlations with the $T_c$ enhancement, although some theoretical calculations propose a $T_c$ peak at a nematic QCP \cite{Lederer2015}. Possible origins of this unexpected result are related to the fact that the nematic order is most likely a ferro-type order with the wave vector ${\bf q}=0$, which is different from AFM order with a finite ${\bf q}$. In a ferro-type order, the Fermi surface change is not as dramatic as in the antiferro-type order in which band-folding occurs, and thus the competition between superconductivity and ferro-nematic order may be much more modest. Indeed, the unusual lack of coupling between superconductivity and orthorhombic distortion in nonmagnetic FeSe has been reported by the thermal expansion measurements \cite{Boehmer2013}. Another factor which may be relevant here is the enhanced quasiparticle damping effect near a ferro-type QCP, that can suppress $T_{\rm c}$. In fact, such a scenario has been discussed in ferromagnetic superconductors, such as UGe$_2$, in which the peak of the $T_{\rm c}$ dome is not located at a ferromagnetic end point \cite{Lohneysen2007}. However, the resistivity curves do not show a dramatic change near the present nematic QCP, and thus further studies are necessary to clarify this point.

It should be noted that in cuprate superconductors the importance of nematic electronic correlations has been suggested as well \cite{Lederer2015,Maier2014}, and the quantum criticality of the CDW or the pseudogap phase has been discussed \cite{Keimer2015}. In YBa$_2$Cu$_3$O$_{7-\delta}$ \cite{Ramshaw2015} and YBa$_2$Cu$_4$O$_8$ \cite{Putzke2016}, however, opposite correlations between $T_{\rm c}$ and the quasiparticle mass enhancement associated with the putative quantum phase transition have been reported, causing controversy on the relationship between quantum fluctuations and superconductivity in cuprates. To solve this issue it would be important to separate the effects of CDW order from the nematic correlations in the pseudogap state. 

The present discovery of the nematic QCP and enhanced ferro-type nematic fluctuations without coexisting or competing magnetic order in FeSe$_{1-x}$S$_x$ may provide a unique avenue to study the unconventional superconductivity mediated by exotic mechanisms other than spin fluctuations.

\section*{Methods} 
Single crystals of FeSe$_{1-x}$S$_x$ were grown by the chemical vapour transport technique. A mixture of Fe, Se and S powders together with KCl and AlCl$_3$ powders was sealed in an evacuated SiO$_2$ tube. The ampoule was heated to $390-450^\circ$C on one end while the other end was kept at $140-200^\circ$C. In these conditions, we were able to cover a much wider $x$ range than previous single crystal studies \cite{Watson2015,Moore2015}, which is important to completely suppress the nematic transition. 
The actual sulphur composition of $x$ is determined by the energy dispersive X-ray spectroscopy, and is found to be about 80\% of the nominal S content (Fig.\;1, inset). The experimental setup for the measurement of nematic susceptibility $\chi_{\rm nem}$ is shown in the inset of Fig.\;2 \cite{Kuo2013}. The samples are cut into bar-shape along the orthogonal crystal axis (tetragonal $[110]$) and are glued on a piezoelectric stack. After mounted on the piezoelectric stack, samples are cleaved so that the thickness becomes about $\sim 30\,\mu$m and the strain is sufficiently transmitted to them. The strain $\epsilon = \frac{\Delta L}{L}$ is induced by applying the voltage to piezoelectric stack and measured by strain gauges glued to the surface of piezoelectric stack. We simultaneously measure the changes in the resistance when varying the strain and this induced anisotropy directly couples to the electronic nematic order parameter. 

\section*{Acknowledgements}
We thank J.-H. Chu, A.\,I. Coldea, R.\,M. Fernandes, I.\,R. Fisher, C.\,W. Hicks, H. Kontani, S. Lederer, A.\,H. Nevidomskyy, T. Takimoto, M.\,D. Watson, and T. Wolf for helpful comments and discussions. This work was supported by Grants-in-Aid for Scientific Research (KAKENHI) from Japan Society for the Promotion of Science (JSPS), and by the `Topological Material Science' Grant-in-Aid for Scientific Research on Innovative Areas from the Ministry of Education, Culture, Sports, Science and Technology (MEXT) of Japan. 


\begin{figure}[t]
	\begin{center}
	\includegraphics[width=0.6\linewidth]{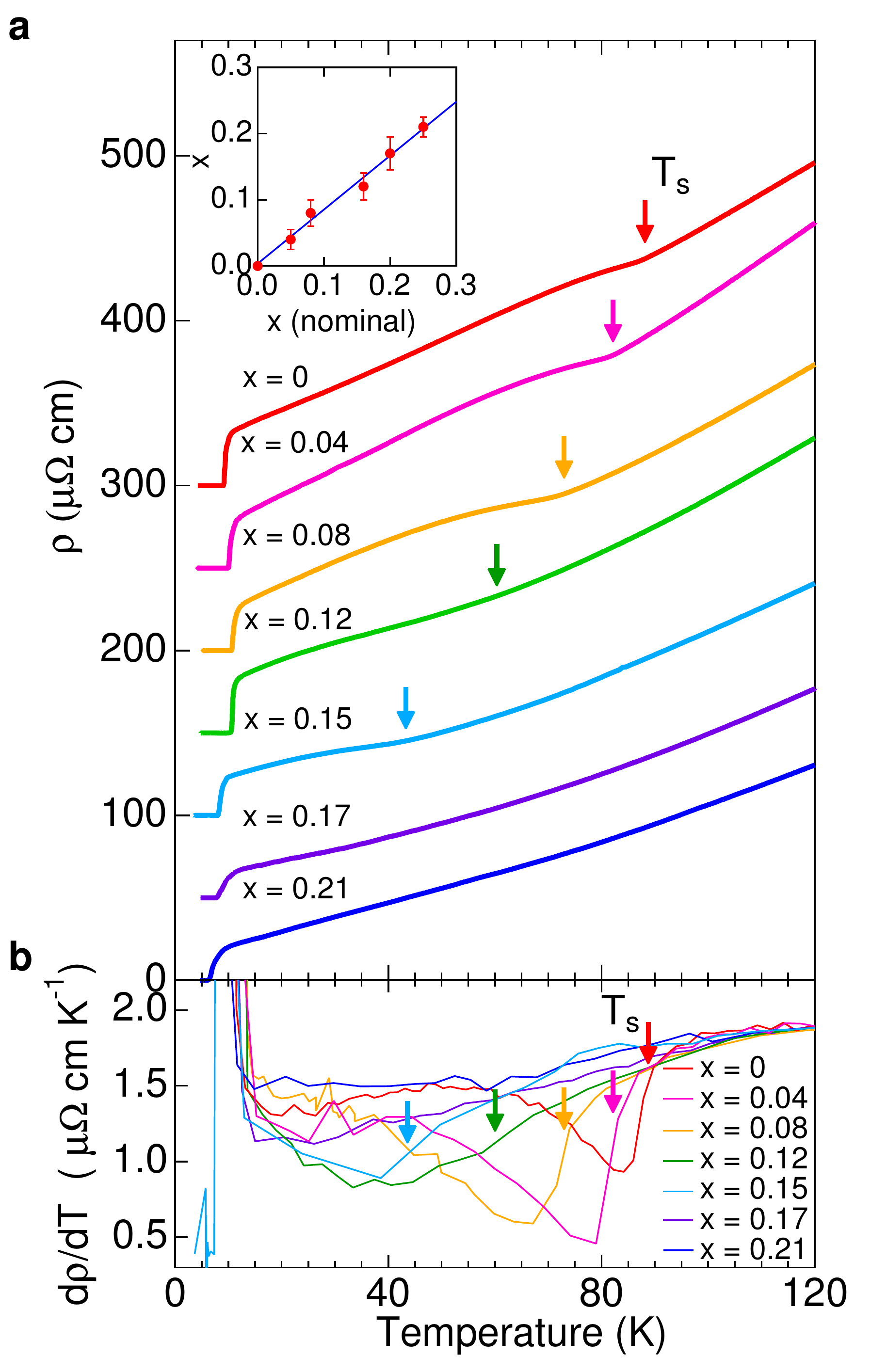}
	\end{center}
	\caption{{\bf Temperature dependence of resistivity in FeSe$_{1-x}$S$_{x}$ single crystals.} {\bf a}, In-plane resistivity $\rho$ as a function of temperature in several crystals with different $x$. Each curve is shifted vertically for clarity. Sulphur content $x$ determined by energy dispersive X-ray spectroscopy is linearly proportional to the nominal composition (inset).  {\bf b}, Temperature dependence of ${\rm d}\rho/{\rm d}T$. Each curve is shifted vertically so that all the data at 120\,K merge. The arrows indicate the nematic (structural) transition temperature $T_{\rm s}$ estimated from the inflection point, which is more accurately determined by the peak position in the temperature dependence of the nematic susceptibility (Figs.\;3a-c).
	}
	\label{fig1}
\end{figure}
  \begin{figure}[t]
	\begin{center}
		\includegraphics[width=0.8\linewidth]{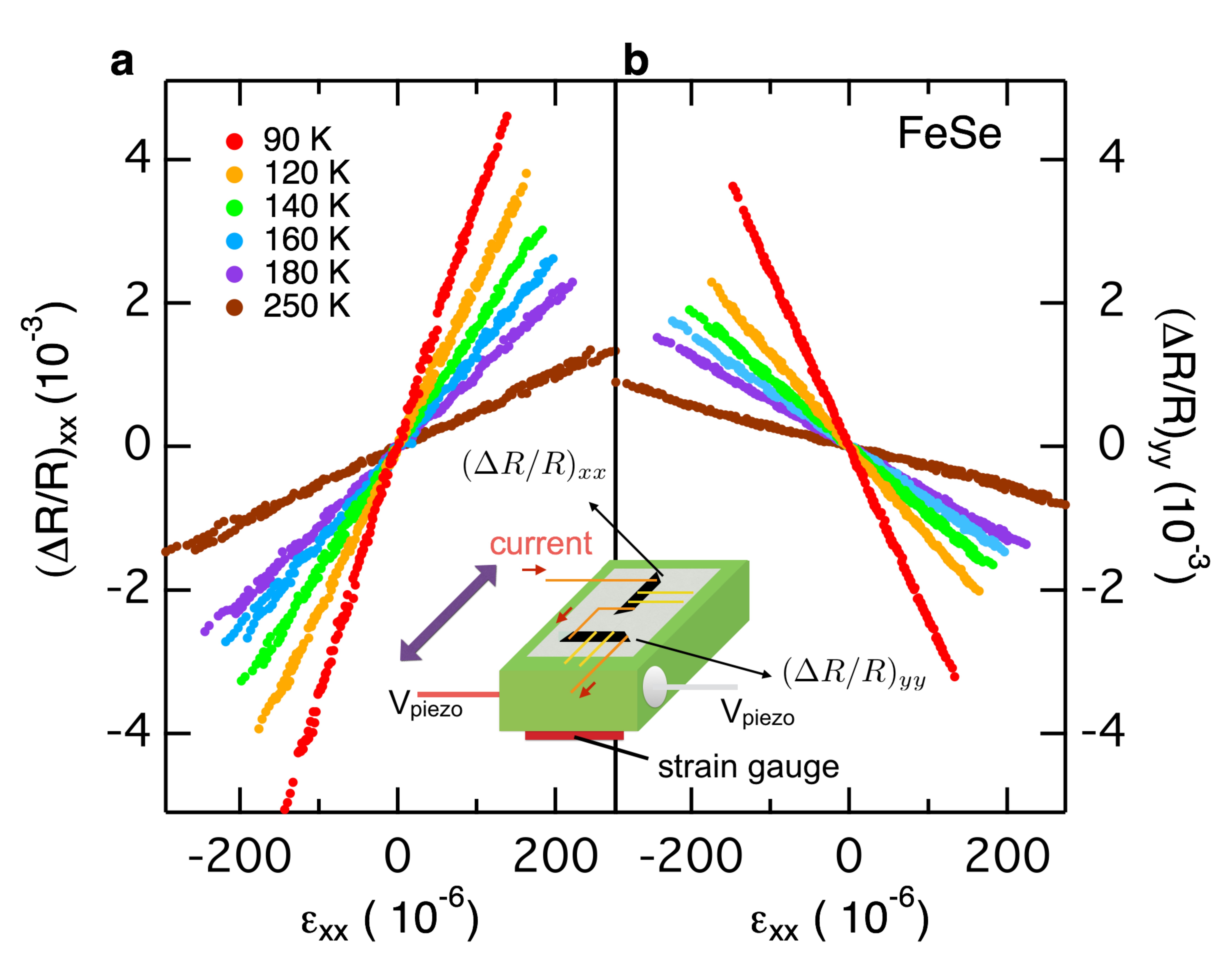}
	\end{center}
\caption{{\bf Elastoresistance measurements by using the piezoelectric device.} {\bf a}, Relative change in the resistance of an FeSe crystal at several temperatures above $T_{\rm s}$ as a function of strain induced by the attached piezoelectric stack. The current direction in the crystal is parallel to the strain direction. {\bf b}, Similar data for another crystal in which the current direction in perpendicular to the strain direction. Inset is a schematic experimental set-up. The crystals (black bars) are attached on top of the piezoelectric stack (green) which expands (shrinks) along the violet arrows when positive (negative) bias is applied. The strain is measured by a strain gauge (red) attached underneath the device. 
}
 \label{fig2}
\end{figure}

  \begin{figure}[t]
	\begin{center}
		\includegraphics[width=\linewidth]{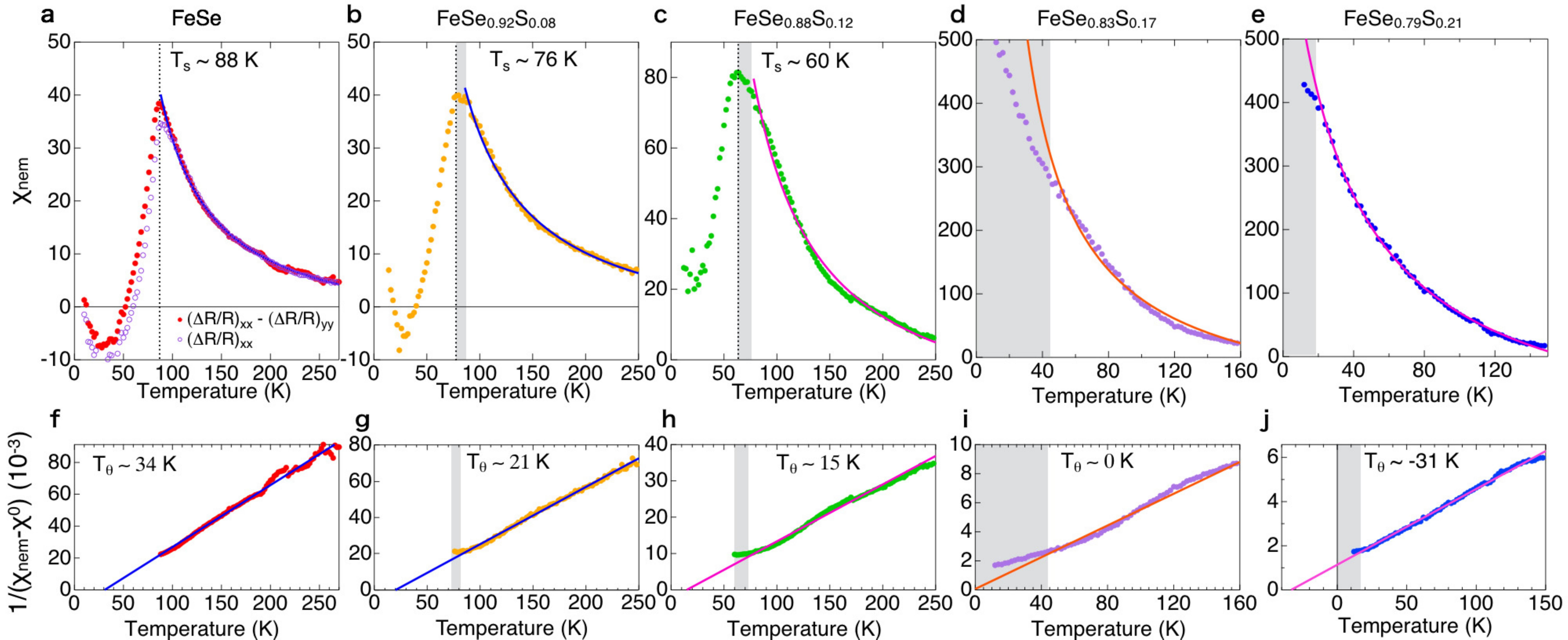}
	\end{center}
\caption{{\bf Divergent nematic susceptibility in FeSe$_{1-x}$S$_{x}$ single crystals.} {\bf a-e}, Temperature dependence of the nematic susceptibility $\chi_{\rm nem}$ for $x=0$ ({\bf a}), 0,08 ({\bf b}),  0.12 ({\bf c}),  0.17({\bf d}), and 0.21 ({\bf e}). In FeSe ($x=0$), the nematic susceptibility data extracted by using both $(\Delta R/R)_{xx}-(\Delta R/R)_{yy}$ (red filled circles) and $(\Delta R/R)_{xx}$ (violet open circles) show consistent results. For other $x>0$ only $(\Delta R/R)_{xx}$ data has been used ({\bf b-e}). {\bf f-j},  Temperature dependence of inverse of  $\chi_{\rm nem}- \chi^{0}$ for $x=0$ ({\bf f}), 0,08 ({\bf g}),  0.12 ({\bf h}),  0.17({\bf i}), and 0.21 ({\bf j}). Here $\chi^{0}$ is a temperature independent constant, coming from the geometrical change that leads to a resistance change under strain. Solid lines are the fits of $\chi_{\rm nem}- \chi_{0}$ to the Curie-Weiss law $\propto 1/(T-T_{\theta})$, and the obtained values of the Weiss temperature $T_\theta$ are indicated in the lower panels. The nematic transition is defined by the peak in $\chi_{\rm nem}(T)$, which is quantitatively consistent with the kink anomaly in $\rho(T)$ curves (Fig.\;1). In the temperature range highlighted in grey, the nematic susceptibility data above $T_{\rm s}$ shows some deviations from the Curie-Weiss law. 
}
 \label{fig3}
\end{figure}

   \begin{figure}[t]
	\begin{center}
		\includegraphics[width=0.7\linewidth]{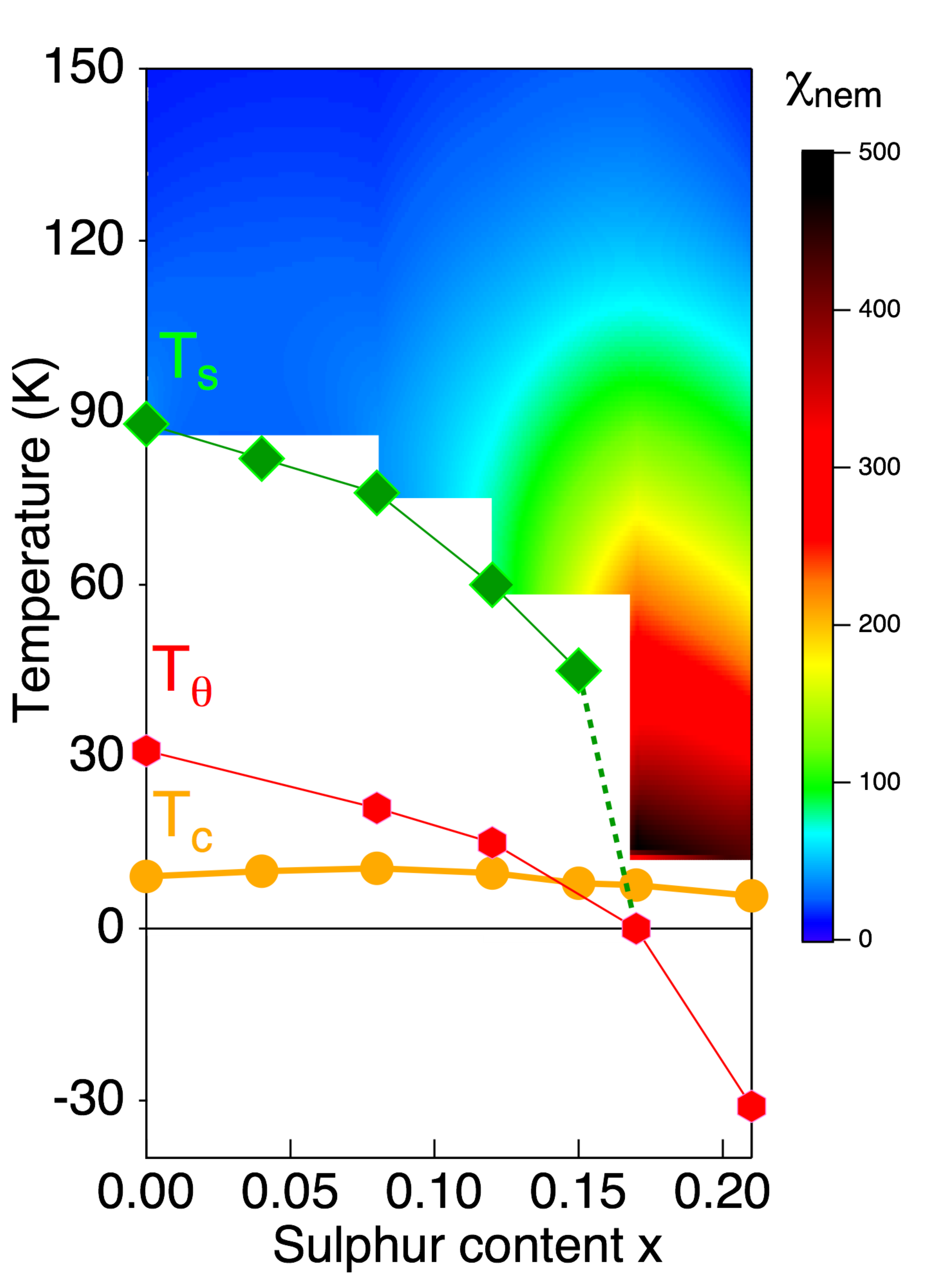}
	\end{center}
   	\caption{{\bf Phase diagram and quantum criticality in FeSe$_{1-x}$S$_{x}$.} Temperature dependence of the nematic transition ($T_{\rm s}$, green diamonds) and the superconducting transition temperature ($T_{\rm c}$, orange circles) determined by the zero resistivity criteria. The Weiss temperature  obtained by the Curie-Weiss analysis of the nematic susceptibility data (Fig.\;3) is also plotted ($T_{\theta}$, red hexagons). The magnitude of $\chi_{\rm nem}$ in the tetragonal phase is superimposed in the phase diagram by a colour contour (see the colour bar for the scale). The lines are the guides for the eyes. 
}
   	\label{fig4}
   \end{figure}

\end{document}